\documentclass[twocolumn,preprintnumbers,amsmath,amssymb]{revtex4}
\usepackage{dcolumn}
\usepackage{color}
\usepackage{bm}
\usepackage{graphicx}
\usepackage{epsfig}
\usepackage{amsmath}
\usepackage{amsfonts}
\usepackage{amssymb}

\def \be{\begin{equation}}
\def \ee{\end{equation}}
\def \beq{\begin{equation}}
\def \eeq{\end{equation}}
\def \bea{\begin{eqnarray}}
\def \eea{\end{eqnarray}}

\def\bfeta{{\boldsymbol{\eta} }}

\def\bA{{\bf A}}

\def\bR{{\bf R}}
\def\bx{{\bf x}}
\def\by{{\bf y}}

\def\br{{\bf r}}
\def\bp{{\bf p}}

\def\bj{{\bf j}}
\def\bk{{\bf k}}

\def\rv{{\rm v}}

\def\cH{{\cal H}}

\def\cA{{\cal A}}

\def\half{{1\over 2}}
\def\vectheta{{\boldsymbol{\Theta}}}

\def\bPi{{\boldsymbol{\Pi}}}

\def\etal{{\it et.~al.~}}
\def\9{\rangle}
\def\6{\langle}
\def\cA{{\cal A}}

\def\sxy{\sigma_{\rm H}}
\begin{document}
\draft

\title{Vortex quantum dynamics of two dimensional  lattice bosons}
\author{Netanel H. Lindner$^{1}$, Assa Auerbach$^{1}$ and Daniel P. Arovas$^{2}$}
\affiliation{1)  Physics Department, Technion, 32000 Haifa,
Israel\\
2) Department of Physics, University of California at San Diego, La
Jolla, CA 92093, USA}

\maketitle
{\bf Vortices, which are introduced into a boson superfluid by
rotation or a magnetic field, tend to localize in a lattice
configuration which coexists with superfluidity
\cite{BEC1,BEC2,BEC3}. In two dimensions a vortex lattice can melt
by quantum  fluctuations resulting in a non-superfluid Quantum
Vortex Liquid (QVL).  Present microscopic understanding of vortex
dynamics of lattice bosons is insufficient to predict the actual
melting density. A missing energy scale, which is difficult to
obtain perturbatively or semiclassically, is the  ``bare'' vortex
hopping rate $t_{\rm v}$ on the dual lattice. Another puzzle is
the temperature dependent Hall conductivity $\sigma_{\rm H}(T)$,
which reflects the effective vortex Magnus dynamics in the QVL
phase. In this paper we compute $t_{\rm v}$ and $\sigma_{\rm
H}(T)$ by exact diagonalization of finite clusters near half
filling. Mapping our effective Hamiltonian  to the Boson Coloumb
Liquid simulated by Ref. \cite{margo}, we expect a QVL above a
melting density of $6.5
\times 10^{-3}$ vortices per lattice site.  The  Hall conductivity
near half filling reverses sign in a sharp transition accompanied
by a vanishing temperature scale. At half filling,  we show that
vortices carry spin half degrees of freedom (`v-spins'), as a
consequence of local non commuting SU(2) symmetries. Our findings
could be realized in cold atoms, Josephson junction arrays and
cuprate superconductors. }

\begin{figure}[htb]
\includegraphics[width=8cm,angle=0]{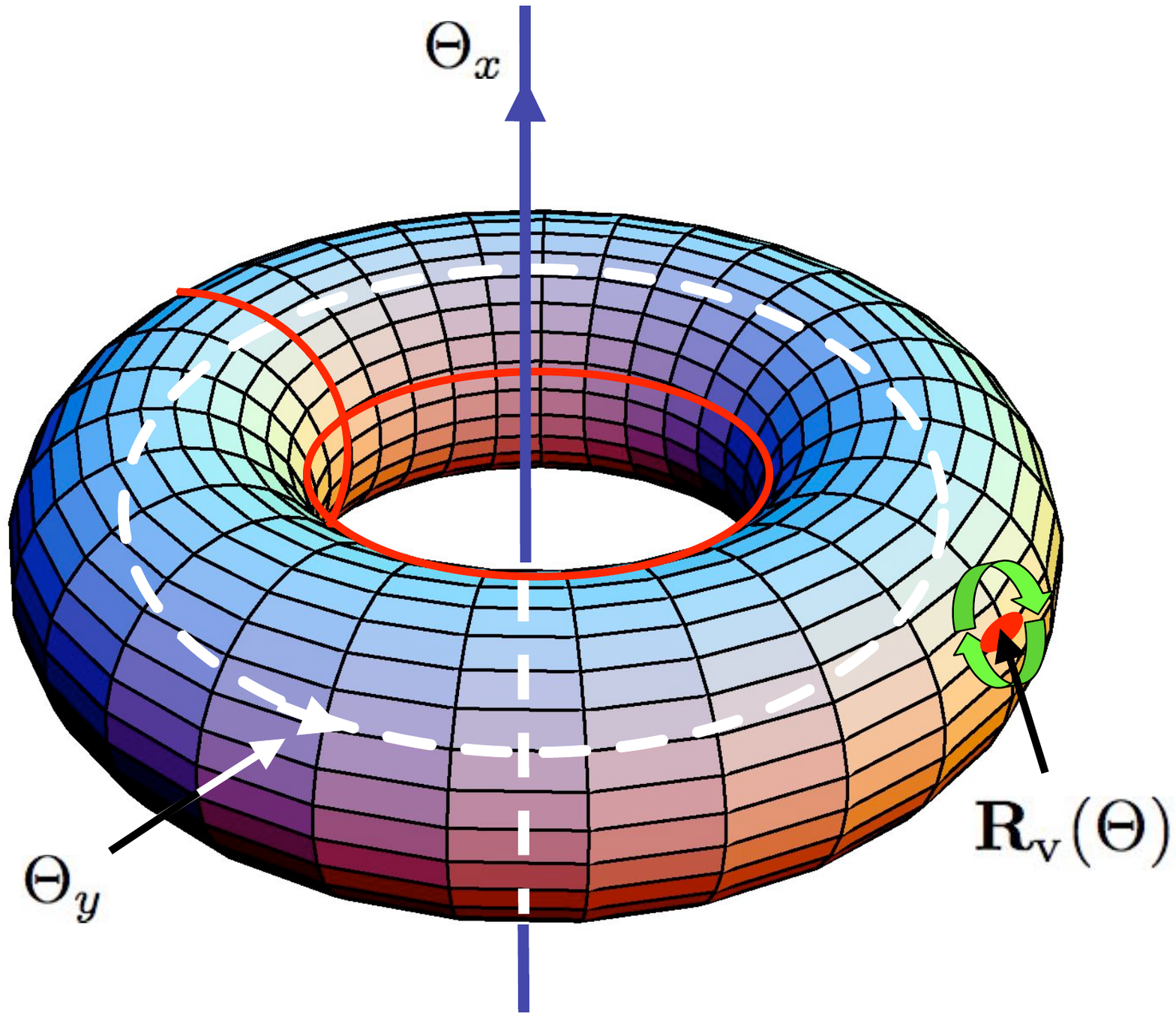}\vspace{-1cm}
\caption{The Gauged Torus.  Geometry of HCB Hamiltonian Eq. (\ref{xxz}) which serves to extract its vortex mass and Hall conductivity.
The torus surface  is penetrated by a uniform magnetic field of one flux quantum, and
threaded by two Aharonov Bohm fluxes
$\vectheta=(\Theta^x,\Theta^y)$. For one flux quantum there is no
translational symmetry on the torus. Red circles denote cycles of
zero flux, and the vortex center $\bR_\rv(\vectheta)$ is localized
on the antipodal point to their intersection, the {\em null
point}. }
\label{fig:torus}
\end{figure}

{\em The model --} We consider $N_b$ hard core bosons (HCB) hopping on a square lattice of unit lattice constant and size $N=L^x  L^y$.
An external vector potential $\bA$ modulates the hopping amplitude  (Josephson energy) $t$.
The system is placed on a torus with periodic boundary conditions, as  shown in Fig. \ref{fig:torus}.

In the spin-$\frac{1}{2}$ representation of HCB, the angular
momentum raising and lowering operators $S^\pm_\br$ create and
annihilate bosons; the occupation number is $n_\br=S^z_\br +\half$.
The Hamiltonian we study is a gauged $XXZ$ model,
\bea
 \cH &=&-{t\over 4}  \sum_{\br,\bfeta}
\left( e^{i  A_\bfeta (\br) } S_{\br}^{+}S^{-}_{\br+\bfeta} +{\rm H.c.}\right)  \nonumber\\
&&\qquad+  {V\over 2} \sum_{\br,\bfeta}
S^z_\br S^z_{\br+\bfeta}  \label{xxz}
\eea
Here $\bfeta=\pm\hat{\bx},\pm\hat{\by}$, is the link direction on which
the lattice gauge field $A_\bfeta$ is defined.  Here we only consider the  superfluid regime of weak nearest
neighbor repulsion $0<V\ll t$.

In the absence of external magnetic
field, the classical ground state of $\cH $, is a ferromagnet in the $XY$ plane
with a uniform $z$-magnetization density $m^z  =   n_b -1/2 $. The
mean field superfluid stiffness is given by $
\rho_s^{\rm mf}= t n_b (1-n_b) $.  Consequently, the superfluid
transition temperature, which is proportional to $\rho_s$, is
maximal at half filling  \cite{comment:MFT}.

An important distinction between lattice hard core bosons and continuum models is
the existence of a charge conjugation operator
$C  \equiv  \exp\left(i\pi  \sum_\br S^x_\br \right)$ on the lattice.  $C$ transforms boson ''particles'' into ''holes''
 $n_i\to (1-n_i)$, and the Hamiltonian into
\be
C^\dagger \cH[\bA, n_b ]C  =   \cH[-\bA, 1-n_b ],
\label{Part-hole}
\ee
where $n_b=N_b/N$  is  the filling fraction.
A consequence of (\ref{Part-hole}) is that the Hall conductivity  is {\em antisymmetric} in  $n_b-1/2$:
\be
\sigma_{H}(n_b,T)  = - \sigma_{H}(1-n_b,T).
\label{phs}
\ee
In terms of vortex motion, this relation implies that below and above half filling vortices drift in opposite
directions relative to the particle current. Sign reversal of  Hall conductivity
 is familiar from tight binding electrons at half filling on bipartite lattices.
Here, however,  the mechanism of sign reversal is different: it arises from the  hard-core interactions
of bosons and occurs for any lattice structure.

{\em Vortex effective Hamiltonian --}  Vortices in a two-dimensional superfluid
act as charges in $(2+1)$-dimensional electrodynamics, where the phonons of the
superfluid become the photons of the electromagnetic theory.  The duality between
vortices and charges is discussed in refs. \cite{POP73,FL89,SIM94,ZEE94,VOL95,ARO97}.
The role of speed of light is played by the phonon speed of
sound, which in our model is  $c= \sqrt{2}t a /\hbar$, where $a$ is the lattice constant.
The vortex centers $\bR_i$  are modeled as point charges hopping
on the dual lattice of plaquette centers, coupled minimally to the
gauge field $\cA^\mu = A^\mu + a^\mu$, where $a^\mu$ is due to the
average boson density, generating a dual magnetic flux per plaquette of $2\pi n_b$,
and where $A^\mu$ is a dynamical field describing the fluctuations in the boson
density and current. At low energies, vortex dynamics is described by a
Harper Hamiltonian plus confining potential,
\beq
H^{\rm v}_{\bR,\bR'} =  -{t_{ v}\over 2} \sum_{ \bfeta } e^{i
\cA^{d}_{\bfeta} }\,\delta_{\bR', \bR+\bfeta}
~+U_N(\bR)~\delta_{\bR ,\bR'}\ .
\label{hb}
\eeq
$U_N(\bR)$ is an effective potential which binds the vortex to the `center of mass' location  $\bR_\rv(\Theta^x, \Theta^y)$.
given by \cite{HR85,ABHLR88},
\be
R^\alpha_{\rm v} ={L^\alpha\over N_\phi}\,\epsilon^{\alpha\beta}
\Big(n_\beta + {\Theta_\beta\over 2\pi}\Big)\mod L^\alpha,
\label{eq:vcenter}
\ee
 $N_\phi$ is the number of flux quanta, $n_\alpha$ are integers,
$L_\alpha$ are the dimensions of the torus.  The phase angles
$\Theta^\alpha$ are proportional to the solenoidal fluxes of the two
elementary toroidal cycles, and a set of `null points' at the intersections
of cycles ${\cal C}_{x,y}$ is defined by requiring $\oint\limits_{{\cal C}_\alpha}
{\bf A}\cdot d{\bf l}=2\pi n_\alpha/N_\phi$, as shown for $N_\phi=1$ in Fig. \ref{fig:torus}.
The vortex CM must lie antipodal to one of the $N_\phi^2$ null points;
we study the simple case $N_\phi=1$.  The constant
$K$ is calculated variationally from Eq.~(\ref{xxz}) using spin
coherent states. Minimizing the energy with respect to the
position of the vortex centered at $\bR$, determines the effective
potential $U_N(\bR)$. For $V=0$ at we find $K
\simeq 39.2\, t n_b(1-n_b)$.  We stress that inclusion of the potential
$U_N(\bR)$ is essential in order to extract the vortex hopping parameters
from our small scale numerical calculations.

For a quantitative quantum theory of vortices we need to evaluate
the effective hopping $t_{\rm v}$. Since vortex tunneling between
lattice sites depend on short range many-body correlations, we
extract $t_{\rm v}$ from exact numerical diagonalizations of $\cH $ on
$16-20$ sites  clusters, in the presence of a single flux quantum.
By tuning $t_{\rm v}$, we fit the lowest three eigenenergies $E_n$ and
eigenstates $|\Psi_n\rangle$ of $\cH$ to the lowest states of the
effective Harper Hamiltonian (\ref{hb}). This assume that
$|\Psi_n\rangle$ states correspond to zero point fluctuations of
the vortex positions, since phonons are frozen out by the finite
lattice gap of order $ 2\pi t /\sqrt{N}$.

\begin{figure}[!t]
\vspace{-0.3cm}
\begin{center}
\includegraphics[width=10cm,angle=0]{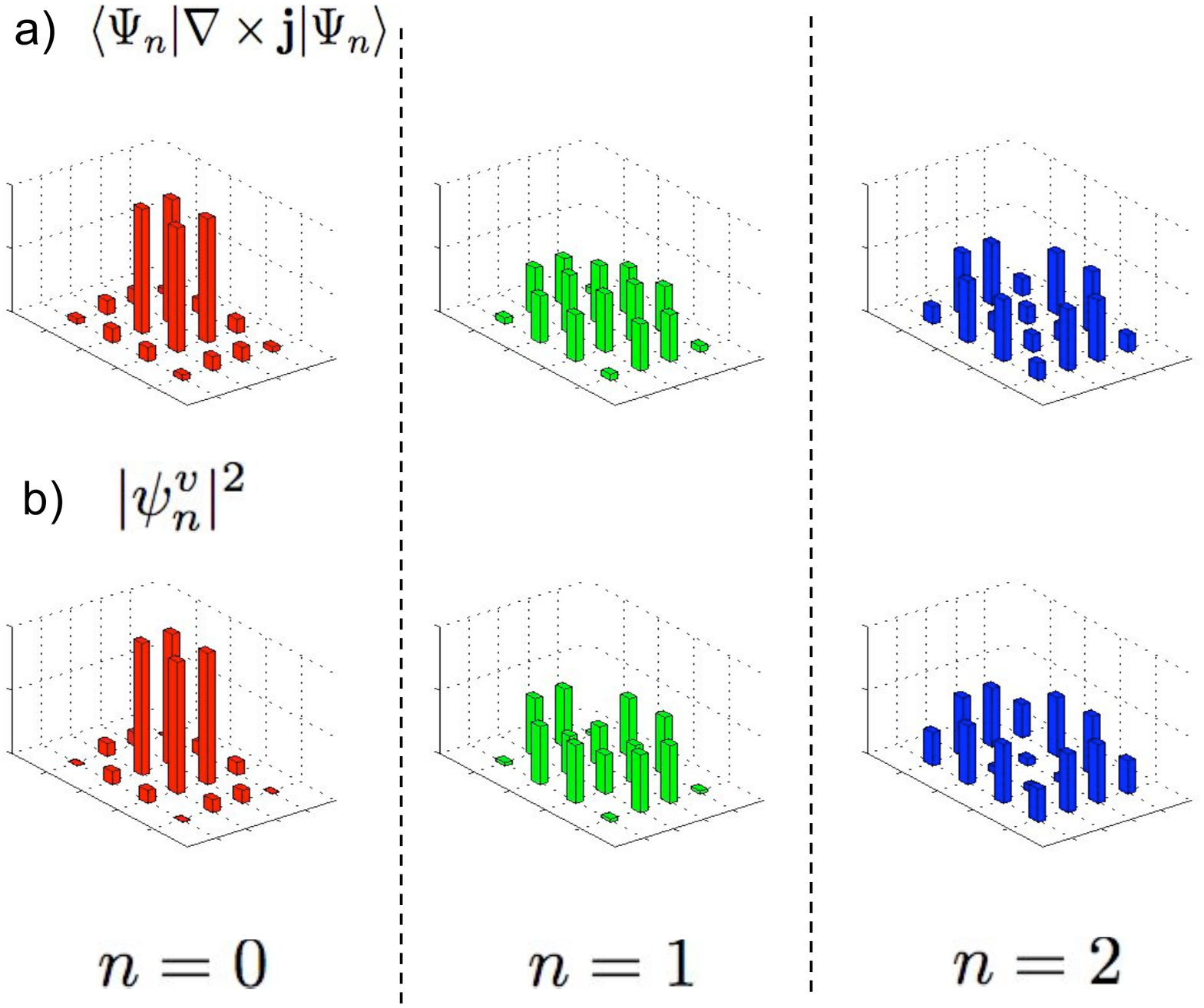}
\vspace{-0.8cm}
\caption{(a) The vorticity $\langle\nabla \times \bj\rangle$ for the first
three doublets of the HCB model, Eq.(\ref{xxz}), with $N_{\phi}=1$  and
$\vectheta=0$ on a $4\times4$ lattice. The uniform background
vorticity has been subtracted. (b) Single particle probability density of the lowest
three excitations of $H^{\rm v}$, Eq. (\ref{hb}) with
$t_{\rm v}=t$.\label{fig:wavefunctions}}

\end{center}
\end{figure}


Our results for $t_{\rm v}(n_b, V/t)$, for $N=20$ fit the analytical
approximations,
\bea
t_{\rm v}  (n_b,0)  &=&  t   - 12.6 \left( n_b-\half\right )^2 + 1264  \left( n_b-\half\right )^4,\nonumber\\
t_{\rm v} (\half,V)  &=&   t  + 1.5 V +2.7 \,{V^2\over t}.
\label{tv}
\eea
The system parameters were varied in the range $|n_b -\half | \le
0.2$, and  $V/t < 0.5$. We find that at half filling,  $t_{\rm v}$
varies very little between the $N=16$ and $N=20$  lattices.
To further test our assignment of $t_{\rm v}$,  we compare the vorticity
density $\langle \nabla\times \bj\rangle$ of eigenstates of $\cH$,
to the probability density of eigenstates of $H^{\rm v}$. As shown in
Fig.~\ref{fig:wavefunctions}, using the fitted value of $t_{\rm v}$ we
obtain similar distributions for both sets of wavefunctions.

{\em Vortex tunneling --} In (\ref{tv}) we find that  near half filling, vortices are as light as bosons,  $t_{\rm v}\approx t$. This implies that the vortex tunneling rate between
two localized pinning potentials of strength $V$, which are separated by distance $d$, decays exponentially as $\Gamma\sim V e^{-d/\lambda}$. The localization length  $\lambda \propto \sqrt{t/V}$ diverges at weak pinning.
This result is to be contrasted with weakly interacting continuum Bose gas. There,
the vortex tunnelling rate between pinning sites is much smaller, and decays as a Gaussian $\Gamma\sim e^{-{\pi\over 2} n_b
d^2}$  \cite{AAG}. From this comparison, we  conclude that at half filling $n_b=1/2$, the lattice and interactions enhance vortex mobility considerably.

{\em Quantum Melting Transition --} Having calculated $t_{\rm v}$, we  can
write down the effective multi-vortex hamiltonian in the
thermodynamic limit. We drop $U_N$ at  large $N$.  By
(\ref{hb}), at half filling  there is dual magnetic flux $\pi$ per
plaquette. In the magnetic Brillouin zone, there is a two-fold
degenerate dispersion $E_{\bk,s}$, $s=\uparrow,\downarrow$. We
later  return to explain the origin of this `v-spin' degeneracy.
The vortex effective mass is $M_{\rm v}^{-1} = \partial^2 E_\bk /
\partial\bk^2 ={t_{\rm v} a^2 /\hbar^2 }$. Integrating out the phonon
fluctuations $A^\mu$,  produces an instantaneous logarithmic (2D
Coulomb) interaction between vortices, plus retarded
(frequency-dependent) interactions \cite{ARO97}. These can be
represented by a self energy $\cH^{\rm ret}(\omega)$. Since we are
interested in the short wavelength fluctuations
which are responsible for the quantum melting of the vortex
lattice, we ignore these retardation effects.

Thus,  for half filled bosons and a vortex density  $n_{\rm v}$ we
arrive multivortex Hamiltonian
\bea
\cH^{\rm mv}  &=& \sum_{i, s=\uparrow\downarrow}    { \bp_i^2 \over  2M_{\rm v}} +  {\pi t\over 4}\sum_{i\ne j}  \log(|\br_i-\br_j|)\nonumber\\
&&~~~~~~ -{n_{\rm v} \pi^2 t\over 4}  \sum_i |\br_i|^2 + \cH^{\rm ret}(\omega).
\label{hamN}
\eea
The single spin version of $\cH^{\rm mv}$, after setting $\cH^{\rm
ret}\to 0$, is the Boson Coloumb Liquid studied by Magro and
Ceperly (MC)
\cite{margo} by diffusion Monte-Carlo simulation. Their
dimensionless parameter which governs the phase diagram is
$r_s^{-2} = \pi n_{\rm v} a_0^2$. We set their $a_0=(\frac{\hbar^2}{\pi
t M_{\rm v}})^{1/2}$ as the microscopic length  which matches between
their model and $\cH^{\rm mv}$.  MC found that below $r_s
\approx 12$ the boson lattice undergoes quantum melting. Using our
values of $t_{\rm v}$ in Eq. (\ref{tv}), the critical $r_s=12$
translates into a  vortex melting  density
 of
 \be
n_{\rm v}^{\rm cr} \le \left(6.5-7.9 {V\over t}\right)\times10^{-3}
~\mbox{vortices per site}.
\label{ncr}
 \ee
This is  a suprisingly low vortex density, which implies that a
QVL can be created at manageable rotation  frequencies for cold
atoms, and moderate magnetic fields for Josephson junction arrays
and cuprate superconductors.


\begin{figure}[!t]
\begin{center}
\includegraphics[width=10cm,angle=0]{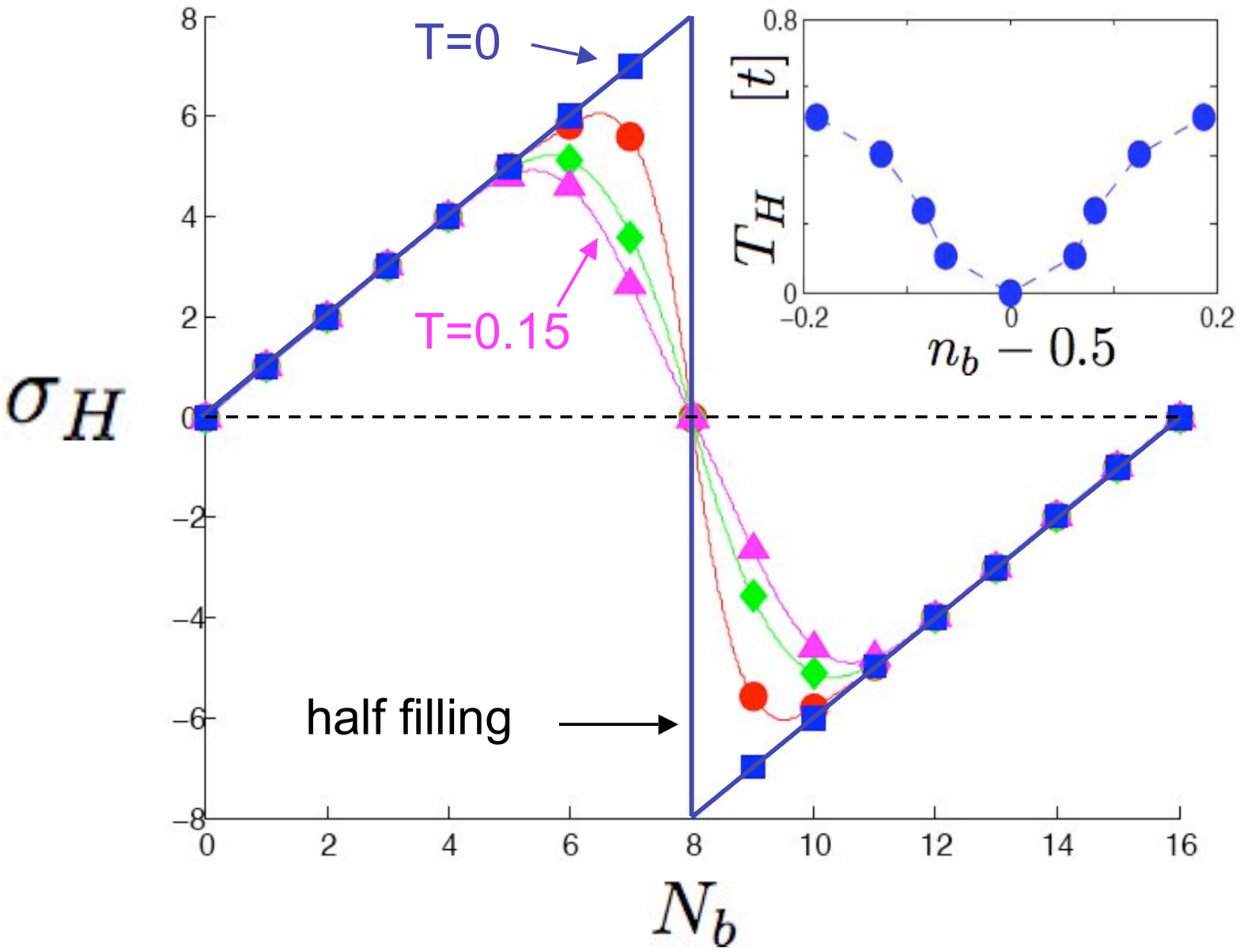}
 \caption{Hall conductance as a function of  boson number $N_b$ for
Hard Core Bosons, Eq. (\ref{xxz}) on the  torus.  Temperatures
vary in intervals of $\Delta T= 0.05 t$. The jump of the zero
temperature conductance at half filling, is smoothened at finite
temperatures. Inset: Hall temperature scale as a function of
density deviation from half filling. $T_{\rm H}$ is defined by
$\sigma_{\rm H}(T_{\rm H})=\half \sigma_{\rm H}(0)$.\label{fig:hall}
 }
\end{center}
\end{figure}


{\em Hall Conductance --}
The temperature-dependent Hall conductance of the finite cluster is
given by the thermally averaged  Chern numbers
\cite{yosi}:
\bea
\sxy(n_b,T) &=&  {1\over \pi  } \sum_{n=0}^{\infty} \int_0^{2\pi}\int_0^{2\pi}\!  d^2 \vectheta\  ~ {e^{- E_n/T }\over Z} \nonumber\\
&&~~~~~~~~~~\times  \mbox{Im}\left\langle{\partial \psi_n \over\partial  \Theta_x}\bigg|
{\partial \psi_n \over\partial  \Theta_y}\right\rangle
\label{sxy-T}
\eea
$E_n(\vectheta)$, and $|\psi_n(\vectheta) \9$  are the exact
spectrum and eigenstates of (\ref{xxz}). The results are matched
at high temperatures with the ones obtained by the Kubo formula
\cite{TBP}.   A typical  Hall conductance as a function of filling for $N_\phi=1$ is plotted in Fig.~\ref{fig:hall}.
At zero temperature, $\sigma_H=N_b$ below half filling.,
reminiscent of the behavior in the continuum $\sxy
\propto N_b/N_{\rm v} $, which holds irrespective of temperatures. However, for HCB $\sxy(T,n_b)$
decreases with temperature. Moreover, $\sxy$ reverses sign at half
filling, as expected by (\ref{phs}).

Our results show a striking general feature. We find that $\sxy$
undergoes a sharp transition between $\sxy>0$ ($\sxy <0$) just
below (above) half filling. As the temperature is lowered, the
sign reversal of the Hall conductance happens across a narrower
region around half filling. This suggests a singularity in the
thermodynamic system with a vanishing energy scale. We define
$T_{\rm H} (n_b)$ by $\sxy(T_{\rm H}) =\half
\sxy(0)$. In the inset of Fig.~\ref{fig:hall}, we show that $T_{\rm H}$
seems to vanish with $|n_b-\half|$, although we cannot yet investigate
this behavior further in larger systems.

{\em  Spin-${1\over 2}$ vortices --}
Half filling is a special density for $\cH$. First, the Hall coefficient  vanishes by (\ref{phs}), which implies that the vortices see no static Magnus field. Second, the external magnetic field  creates a multitude of doublet degeneracies.
 To be precise, for any odd number $N_\phi$  of flux quanta, there are $N$ (the system size) distinct values of AB fluxes $\vectheta_i$  where all  eigenstates are  two-fold degenerate. We have found that these degeneracies are associated with
  non-commuting local symmetry operators
 \be
\Pi^\alpha = \half U^\alpha C P^\alpha[\bR_\rv],~~~\alpha=x,y.
\label{Pialpha}
\ee
$C$ is  the charge conjugation  (see Eq. (\ref{phs})), and $U^\alpha$ is a pure gauge transformation.
$P^{x(y)}$ is a lattice reflection about the $x(y)$ axis passing through the vorticity center $\bR_\rv$. For $N$ discrete AB fluxes, $\bR_{\rm v}$ can be placed on each one of the lattice positions,
where  $[\cH, \Pi^{\alpha}]=0$.
These symmetries  follow from the fact that $C P^\alpha$ preserve  the  magnetic field  and the AB fluxes. If $\vectheta_i$
are tuned by (\ref{eq:vcenter}) to position $\bR_{\rm v}$ on a lattice site,   $\Pi^\alpha $ sends $\cH$ to itself upto   a pure gauge transformation  $\left(U^\alpha\right)^\dagger$.

A straightforward, though cumbersome, calculation \cite{TBP}  yields the
commutation rule
\be
\Pi^y  \Pi^x =(-1)^{N_\phi} \Pi^x \Pi^y . \label{com}
\ee
We  define the vector $\bPi=(\Pi^x,\Pi^y,\Pi^z)$, where  $\Pi^z=2i
\Pi^x\Pi^y$. For any odd number $N_\phi$ of vortices it is easy to
show using (\ref{com}) that each of the energy eigenstates is at
least two-fold degenerate.

Since $ \bPi^2 = 3/4$, they obey the algebra of  spin half
operators.  Thus the doublets reflect the Kramers degeneracy
expected for an odd number of interacting spin half degrees of
freedom which we label  {\em v-spins}. As shown by the form of
$\Pi^\alpha$, the v-spins are attached to the vortex positions.
The $z$-direction polarization corresponds to a boson charge
density wave (CDW) modulation. Variational calculation shows the
CDW to be exponentially localized in the vortex core
\cite{SCvortex}. Thus v-spin interactions between different vortices
decay exponentially, and are very weak in the vortex lattice
regime. We cannot determine their ordering configuration. If they
interact ferromagnetically, a CDW order parameter can form.  In
any case, whatever the ordering tendencies, the low exchange
energy ensures that   v-spins excitations play an important role
in the low temperature thermodynamics and transport coefficients
of the multi vortex system in both lattice and QVL phases.

{\em Nature of the QVL --} Theoretical treatments of lattice
bosons have found  a myriad of vortex-antivortex condensate (VC) phases
at all rational boson filling fractions, $n_b=p/q$, due to
$q$-fold degeneracies of the Harper hamiltonian on an infinite
lattice
\cite{Lannert,Subir,Subir2,Zlatko1,Zlatko2}. VC's  are  in effect, insulating phases where the dual
Anderson-Higgs mechanism produces a Mott gap \cite{FL89}. Some of
these phases result in $q$-periodic CDWs. Therefore a possible
candidate for the VC  at half filling  is the bipartite CDW i.e.
the  antiferromagnetic Ising state of (\ref{xxz}) at $V>t$.
However, the QVL we study, which contains a net density of
vortices, differs from the proposed VC phases in two important
respects.

(i)  MC  \cite{margo} have found  that the liquid phase of $\cH^{\rm mv}$   (\ref{hamN})
has vanishing condensate fraction. If their results (ignoring retardation effects) is relevant to the QVL,
it should differ  from a  charge-gapped
insulator. Whether it is a  metal is an open possibility. Away
from half filling, our results for  $\sigma_{xy}$ show that the
vortices are subject to a strong magnetic field, which further
suppresses their condensation. At low boson fillings and large
vortex density, $n_b/n_{\phi}<1$, there is  evidence for
fractional quantum hall phases \cite{sorensen, demler}.

(ii) Away from  the commensurate filling $n_b=p/q$, the Hall
conductivity is expected to cross zero, and be proportional to the
excess density from $p/q$.  We found that the Hall conductance has
a very different behavior: it has only one abrupt jump between a
positive value below, and negative above half filling.  We have
computed the single flux Hall conductance on a finite lattice, but
we expect it  to represent  the macroscopic Hall conductivity in
the QVL phase where superfluid order parameter correlations are
short ranged.

{\em Summary --}
Vortices of hard core bosons near half filling are highly mobile
logarithmically interacting charges.  The vortex  lattice is expected to melt into a QVL at around
$7\times10^{-3}$ vortices per site. At boson density of half filling,  the vortices
are not subjected to an effective Magnus field, but  carry
local v-spin degrees of freedom, which effect the low energy
correlations. Away from half filling, the Hall conductivity exhibits rapid variation
accompanied by a vanishing energy scale.
Although the issue is far from settled, we present arguments
that the QVL is not a vortex condensate.

{\bf Acknowledgements.}
We thank  Ehud Altman,  Yosi Avron, David
Ceperley,  Herb Fertig, Gil Refael and Efrat Shimshoni  for useful
discussions.  Support of the US Israel Binational Science
Foundation  is gratefully acknowledged.  AA acknowledges the Aspen
Center For Physics where some of the ideas were conceived.

\end{document}